\documentclass[journal=nalefd,manuscript=letter]{achemso}

\usepackage{upgreek}
\usepackage[version=3]{mhchem} % Formula subscripts using \ce{}
\usepackage[T1]{fontenc}       % Use modern font encodings
\usepackage{color}
\usepackage[normalem]{ulem}
\graphicspath{{figures/}}

\author{\'Aron Szab\'o$^{\dagger}$}
\author{Achint Jain$^{\ddagger}$}
\author{Markus Parzefall$^{\ddagger}$}
\author{Lukas Novotny$^{\ddagger}$}
\author{Mathieu Luisier$^{\dagger}$}
\email{mluisier@iis.ee.ethz.ch}
\phone{+41 44 632 53 33}
\fax{+41 44 632 11 94}
\affiliation{$^{\dagger}$ Integrated System Laboratory, ETH Z\"urich,
  8092 Z\"urich, Switzerland\\ $^{\ddagger}$ Photonics Laboratory, ETH
  Z\"urich, 8093 Z\"urich, Switzerland}

\title{Electron transport through metal/MoS$_2$ interfaces: edge- or
  area-dependent process?}

\begin{document}

\begin{abstract}
In ultra-thin two-dimensional (2-D) materials, the formation of ohmic
contacts with top metallic layers is a challenging task that involves
different processes than in bulk-like structures. Besides the Schottky
barrier height, the transfer length of electrons between metals and
2-D monolayers is a highly relevant parameter. For MoS$_2$, both short
($\le$30 nm) and long ($\ge$0.5 $\upmu$m) values have been reported,
corresponding to either an abrupt carrier injection at the contact
edge or a more gradual transfer of electrons over a large contact
area. Here we use \textit{ab initio} quantum transport simulations
to demonstrate that the presence of an oxide layer between a metallic
contact and a MoS$_2$ monolayer, for example TiO$_2$ in case of
titanium electrodes, favors an area-dependent process with a long
transfer length, while a perfectly clean metal-semiconductor interface
would lead to an edge process. These findings reconcile several
theories that have been postulated about the physics of metal/MoS$_2$
interfaces and provide a framework to design future devices with lower
contact resistances.
 
\end{abstract}

Transistors made of novel two-dimensional (2-D) materials beyond
graphene such as single-layer MoS$_2$ \cite{single_layer_mos2} have
generated considerable excitement among the scientific community for
their potential as active components of future integrated
circuits. Transition metal dichalcogenides (TMDs) \cite{TMD}, black
phosphorus \cite{BP_anisotropy,BP_ye}, and hundreds of other presumably
exfoliable 2-D monolayers \cite{high_throughput} appear as excellent
candidates to outperform Si FinFETs, the current workhorse of the
semiconductor industry, for next-generation ultra-scaled logic 
switches \cite{kuhn}. The advantages of 2-D materials over competing
technologies reside in their naturally passivated surfaces, their
planar geometry providing an excellent electrostatic control
\cite{fiori}, their exceptionally high carrier mobilities as compared
to 3-D compounds with the same sub-nanometer thickness
\cite{mob1,mob2,mob3}, and the possibility of stacking them on 
top of each other to form van der Waals heterostructures
\cite{geim,vdw,ianna}.

Before MoS$_2$ field-effect transistors (FETs) with a monolayer
channel can reach their full potential and deliver the expected
performance \cite{yoon}, several key challenges remain to be
solved. The source and drain contact resistances represent one of the
main limiting factors as they usually lie in the k$\Omega\cdot\upmu$m
range \cite{contact,franklin}, instead of 150 to 200 $\Omega\cdot\upmu$m
as in conventional Si transistors \footnote{https://irds.ieee.org/roadmap-2017}. 
Lower values have been reported with metalized 1T MoS$_2$ \cite{1T} or
nickel-etched graphene \cite{ni} electrodes, in the order of 200
$\Omega\cdot\upmu$m, but for multilayer MoS$_2$. While top contacts are
the most widely used variants due to their ease of fabrication, side
contacts have started to emerge as a promising alternative
\cite{side1,side2,side3,side4}, motivated by theoretical studies that predict a
stronger orbital overlap and shorter tunneling distances between
metals and MoS$_2$ in lateral configurations \cite{Kang,Robertson}. 
Apart from the electrode geometry, other well-known techniques have
been applied to reduce the contact resistance of MoS$_2$ FETs, among
them the usage of different metals \cite{das,banerjee}, the
introduction of an interfacial layer between the metal and
semiconductor \cite{kim,lodha,cho}, or the doping of MoS$_2$
\cite{seabaugh,wallace}. Despite significant progresses made over the
last few years, metal/MoS$_2$ interfaces have not yet revealed all their
secrets, hindering the development of future electronic components
based on 2-D materials.

\begin{figure}[t]
 \includegraphics[width=0.5\linewidth]{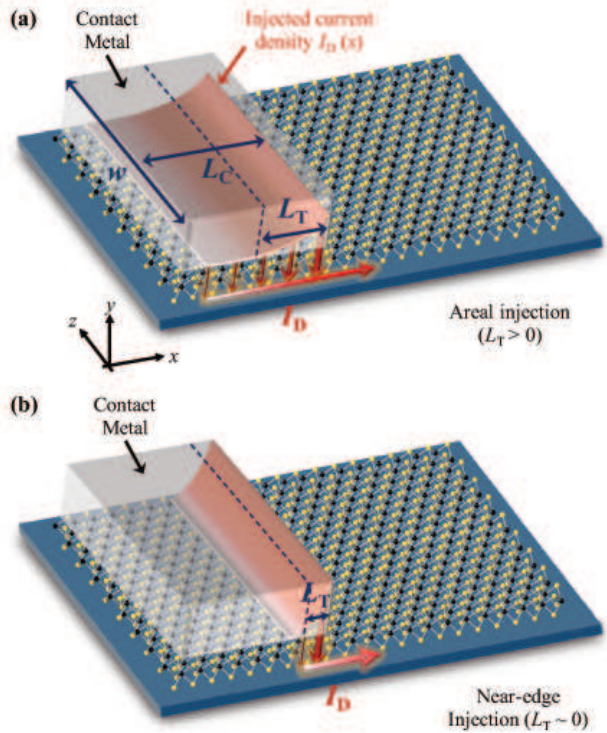}
  \caption{Schematics of top metallic electrodes (gray blocks of width
    $w$ and length $L_C$) deposited on 2-D monolayers (atomic
    structures) with two possible electron injection mechanisms. The
    behavior of the current density $I_D$ flowing through these
    heterojunctions is represented by the red surface plots and
    vertical arrows. (a) Area-dependent injection with a long transfer
    length $L_T$. The amount of current penetrating into the 2-D
    material gradually increases over a metal-semiconductor overlap
    distance equivalent to $L_T$ along the $x$-axis (transport
    direction). (b) Near-edge injection with a close to zero
    $L_T$. Almost all electrons are transferred from the contact to
    the 2-D materials at the edge of the overlap region.} 
  \label{fig:fig0}
\end{figure}

An open issue of critical importance concerns the trajectories
followed by electrons leaving a top metal contact and entering a
MoS$_2$ monolayer situated underneath. The transfer length $L_T$, as
illustrated in Fig.~\ref{fig:fig0}, characterizes the average distance
that is needed by carriers to accomplish this transition. No consensus
exists on the magnitude of $L_T$, i.e. on whether the electrical
current flows through the metal up to the edge of its interface with
MoS$_2$ (edge process, $L_T\simeq0$) or whether electrons
gradually penetrate into MoS$_2$ over a relatively long distance
(area-dependent process, large $L_T$). In the former case, the current
would be proportional to the width $w$ of the electrode on MoS$_2$,
whereas in the latter, it would also depend on the contact length
$L_C$, i.e. on the length of the metal/MoS$_2$ overlap region,
provided that $L_C<L_T$. Experimentally, Liu et
al. \cite{large_transfer_length} deduced a transfer length of about
600 nm for titanium-connected single-layer MoS$_2$. Meanwhile, English
et al. \cite{English} reported pure Au electrodes on top of bilayer
MoS$_2$ with low contact resistances $R_C$=740 $\Omega\cdot\upmu$m and
$L_T\simeq$30 nm, which can be considered a near-edge process. 

Obviously, these results are in total contradiction. Various modeling
efforts have attempted to identify transfer mechanisms that
could explain these opposite trends, but so far without much
success. A prominent study relying on density functional theory (DFT),
an \textit{ab initio} method, concluded that the transfer of  
electrons from Ti to MoS$_2$ should be area-dependent as this 2-D
material gets metalized when put in contact with titanium
\cite{Kang}. Another work combining DFT and quantum transport
calculations within the framework of the Non-equilibrium Green's
Function (NEGF) formalism found that carriers preferably escape the
Ti electrode at the edge of the Ti/MoS$_2$ interface
\cite{iedm_contacts}. Finally, device simulations performed in the 
effective mass approximation suggested that the transfer length
increases with the MoS$_2$ thickness, going from an edge process in
monolayers to an area-dependent injection in multi-layer crystals
\cite{meff_contact}. The apparently contradictory conclusions of these
theoretical investigations do not resolve the discrepancies observed
experimentally. 

To address them, a deeper look at the fabrication of top contacts on
2-D materials should be taken. It has been recently shown that if
titanium is deposited on MoS$_2$ under moderately high vacuum
conditions ($\sim$10$^{-6}$ mbar), a TiO$_2$ oxide layer forms at the
metal-semiconductor interface. Under ultra-high vacuum
($\sim$10$^{-9}$ mbar), this does not happen, allowing the Ti
atoms to directly bind to the top S atoms \cite{mcdonnel}. In this
context, the main difference between the results of
Refs.~\cite{large_transfer_length} and \cite{English} may be
the pressure at which the contacts were deposited (moderately high vs. ultra-high
vacuum), not the choice of the electrode metal (Ti vs. Au) or the
MoS$_2$ thickness (mono- vs. bi-layer). It is likely that an
interfacial oxide layer was present within the Ti/MoS$_2$ stacks of
Ref.~\cite{large_transfer_length}, but not within the Au/MoS$_2$ ones
of Ref.~\cite{English}. The goal of this paper is therefore to analyze
the role played by such layers in the contact physics, to determine
their influence on the transfer length of electrons at metal/MoS$_2$
junctions, and to find out whether they can be leveraged to reduce the
contact resistance of 2-D devices. To do that, we performed \textit{ab
  initio} quantum transport simulations and demonstrated that in the
absence of an intercalated oxide layer between the top metal contact
and MoS$_2$, the transfer of electrons becomes
edge-dependent. Finally, the aforementioned results from the
literature are re-examined in light of these findings.  

\begin{figure}[t]
 \includegraphics[width=\linewidth]{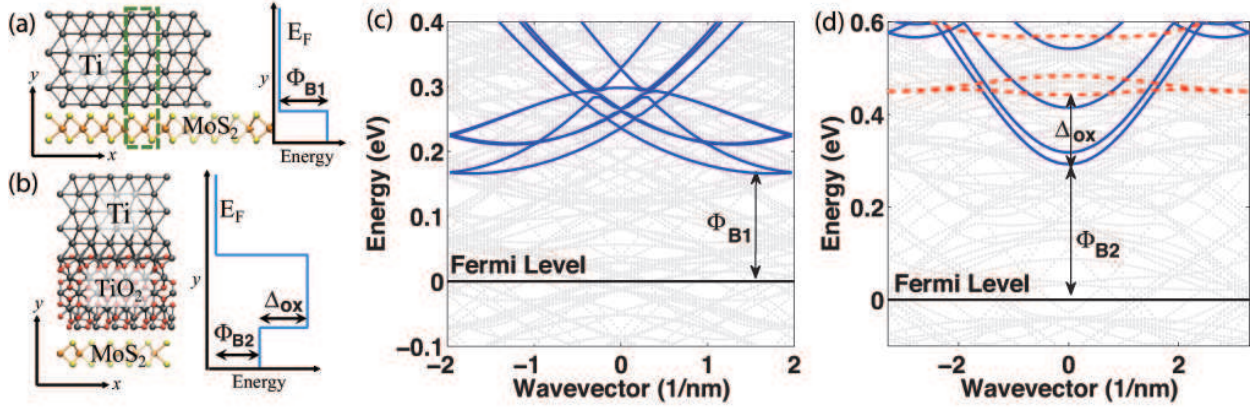}
  \caption{Schematic view of the atomic unit cells of the (a)
    Ti-MoS$_2$ and (b) Ti-TiO$_2$-MoS$_2$ contact geometries simulated
    with DFT and used to construct MLWF-based Hamiltonian matrices. The
    colored spheres represent individual atoms: gray - Ti, red - O,
    orange - Mo, and yellow - S. The right sub-plots show the
    corresponding band alignments. In (a), only a small portion of the 
    unit cell is depicted as the real one contains 576 atoms, whereas
    in (b), the 306 considered atoms are plotted. (c) Electronic
    bandstructure around the Fermi level $E_f$=0 eV for the Ti-MoS$_2$
    system. The primitive unit cell delimited by the dashed green
    rectangle in (a) served as input to this calculation. The dotted
    gray lines are bands lying within the Ti contacts, blue lines within
    MoS$_2$. A Schottky barrier height (SBH) $\Phi_{B1}$=166 meV can
    be extracted. (d) Same as (c), but for the Ti-TiO$_2$-MoS$_2$ unit
    cell in (b). The dashed red lines refer to the TiO$_2$
    bands. Here, a SBH $\Phi_{B2}$=293 meV was found. The
    conduction band offset between MoS$_2$ and TiO$_2$ was adjusted to
    $\Delta_{ox}$=150 meV. Because of the different shape of the
    MoS$_2$ supercells in (a) and (b), the conduction band minimum of
    this material in (c) and (d) was not folded to the same $k$-point.} 
  \label{fig:fig2}
\end{figure}

To highlight the impact of an interfacial TiO$_2$ oxide layer on the
electron transfer process, we simulated Ti-contacted single-layer
MoS$_2$ structures, with and without TiO$_2$ in between, by combining
plane-wave DFT, maximally localized Wannier functions (MLWFs), and NEGF
\cite{prb_szabo}. The first step consisted of creating
suitable atomic geometries. Since DFT calculations are computationally
very demanding, periodic unit cells of the smallest possible
dimensions were created, while still large enough to allow for the
extraction of the bandstructure and wavefunction of three distinct
regions: one with pure Ti, another one made of Ti-(TiO$_2$)-MoS$_2$,
and a last one containing MoS$_2$ only. For simplicity,
  the underlying substrate layer was not included in the DFT study.
Atomic configurations fulfilling these conditions are illustrated in
Figs.~\ref{fig:fig2}(a) and (b). Next, their electronic structure was
calculated with the VASP \cite{vasp} DFT tool within the generalized
gradient approximation (GGA) of Perdew, Burke, and Ernzerhof
\cite{PBE}. The single-particle wavefunctions were then transformed
into a set of MLWFs with the help of the Wannier90 \cite{wannier90}
package. 

From the chosen unit cells and the produced MLWF Hamiltonian blocks,
larger structures with a $\sim$50~nm long free standing MoS$_2$ part
and metal-semiconductor overlap lengths ranging from 6 to 133~nm
were constructed following the procedure described in Ref.~\cite{mythesis}. 
Even though there is no region with pure Ti in Fig.~\ref{fig:fig2}(a) and
(b), the influence of MoS$_2$ and TiO$_2$/MoS$_2$ on Ti was found
sufficiently weak such that a representative Ti Hamiltonian block
could be safely cut off from the generated heterostructures. Similarly, 
in Fig.~\ref{fig:fig2}(b), the penetration of the metal wavefunction
into the MoS$_2$ monolayer was small enough so that the required
Hamiltonian blocks of free-standing MoS$_2$ could be directly derived
from the matrix elements of the whole stack. It has been previously
verified that the bandstructures of all individual layers (Ti,
TiO$_2$, and MoS$_2$), computed after the MLWF transformation and the
single block extraction, agree well with what is obtained from pure
Ti, TiO$_2$, and MoS$_2$ unit cells \cite{mythesis}. The results are
shown in Figs.~\ref{fig:fig2}(c) and (d).

Although DFT is considered a very accurate technique to capture 
both the conduction (CB) and valence band (VB) states of semiconductors 
and insulators, it suffers from a well-known band gap underestimation
problem. Consequently, the calculated band alignments may not always
be reliable. Furthermore, finite size effects may take place in small
simulation domains, especially with a random placement of the
constituting atoms. This issue was avoided in our DFT calculations,
which are limited in size, by replacing amorphous TiO$_2$ with its
rutile, well-ordered phase. To compensate for this idealization, the
TiO$_2$ CB edge had to be manually raised to 0.15 eV above the MoS$_2$
one, which corresponds to the experimentally determined band offset
\cite{mcdonnel}. The adjusted band diagram is presented in
Fig.~\ref{fig:fig2}, where it can also be seen that the non-altered
Schottky barrier heights, with (0.29 eV) and without (0.17 eV)
TiO$_2$, are close to the experimental ones (0.23 eV) \cite{sbh}.  

\begin{figure}[t]
 \includegraphics[width=0.92\linewidth]{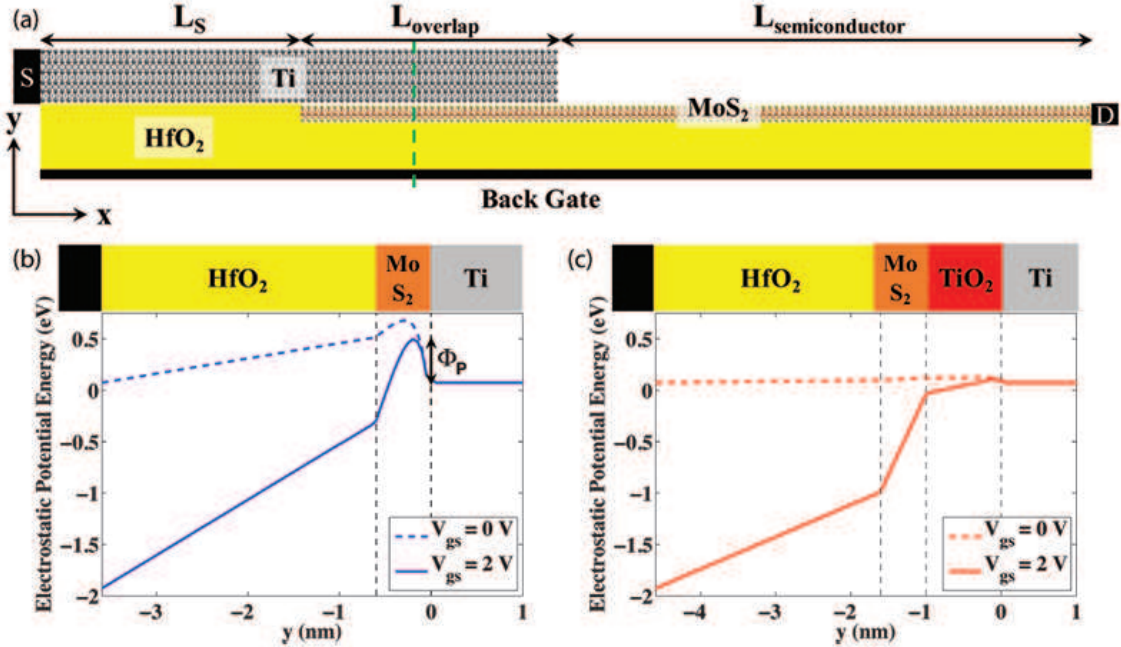}
  \caption{(a) Schematic view of the structures simulated in this
    work. Ti-(TiO$_2$)-MoS$_2$ systems were constructed with a pure Ti
    region of length $L_S$=9.5 nm, a Ti-(TiO$_2$)-MoS$_2$ overlap
    region of length 5.7$\leq L_{overlap} \leq$132.8 nm, and a
    MoS$_2$-only region of length $L_{semiconductor}$=47.7
    nm. Electrons are injected at the contact labeled $S$ (source)
    and collected at the one denoted $D$ (drain). The channel is
    separated from a back gate electrode by a perfectly isolating
    HfO$_2$ oxide layer of thickness $t_{ox}$=3 nm and
      relative dielectric constant $\epsilon_R$=20. Transport occurs
    in the $x$ and $y$ directions, the $z$-axis (out-of-plane) is
    assumed periodic. (b) Electrostatic potential energy along the
    vertical dashed green line in (a) at a back gate voltage
    $V_{gs}$=0 (dashed line) and 2 V (solid line) for the Ti-MoS$_2$
    contact geometry. The variable $\Phi_P$ refers to the
    bias-dependent potential barrier induced by the Ti contact. (c)
    Same as (b), but for the Ti-TiO$_2$-MoS$_2$ configuration. The
    TiO$_2$ layer measures 1 nm in all cases, a value
      large enough to capture the relevant physics, but thin enough to
      remain computationally affordable.} 
  \label{fig:fig3}
\end{figure}

After scaling up the Ti-MoS$_2$ and Ti-TiO$_2$-MoS$_2$ unit cells,
contact geometries similar to the one shown in Fig.~\ref{fig:fig3}(a)
were built up and a 3nm thick HfO$_2$ layer was added to serve as back
gate dielectric. The MLWF Hamiltonian matrices of theses systems were
inserted into a dissipative NEGF quantum transport solver that returns
the current as a function of the applied voltages for a given
electrostatic potential \cite{prb_szabo}. The Green's Functions must
be evaluated for all possible electron energy $E$ and momentum $k_z$
pairs before summing up these contributions to give rise to the
electron density. To reduce the heavy computational burden associated
with NEGF simulations expressed in a MLWF basis, the process was
simplified by pre-computing the electrostatic potential with a
properly calibrated quantum mechanical solver based on the effective
mass picture \cite{ema} and by including only one momentum point
carrying a representative current density. These approximations,
together with the TiO$_2$ band edge shift, might limit the accuracy of
the computed data. Since we focus on a qualitative description of the 
contact physics, the conclusions of the paper are not affected. More
details about the simulation approach can be found in the Supporting
Information. 

Vertical cuts of the electrostatic potential energy across the 
Ti-MoS$_2$ and Ti-TiO$_2$-MoS$_2$ heterojunctions are depicted in
Figs.~\ref{fig:fig3}(b) and (c), respectively, at two different back
gate voltages $V_{gs}$=0 and 2 V. Of particular interest is the
response of the MoS$_2$ monolayer. Without TiO$_2$, a potential
barrier $\Phi_P$ of about 0.5 to 0.6 eV forms at the 
Ti/MoS$_2$ interface, on top of the already existing Schottky barrier
$\Phi_{B1}$=0.166 eV (see Fig.~\ref{fig:fig2}). The origin of $\Phi_P$ 
can be traced back to the penetration of the Ti wavefunctions into the
MoS$_2$ band gap, which induces an additional electron
density of $\sim$3e14 cm$^{-2}$ in MoS$_2$. In Ref.~\cite{Kang} it
was inferred that this phenomenon leads to a metallization of MoS$_2$
and that the created states can carry current, thus causing an
area-dependent injection of electrons below the Ti
contact. It should however be noted that the electrons
  within the MoS$_2$ band gap result from exponentially decaying
  wavefunctions, with a large imaginary component of their wavevector
  along all directions. As these carriers cannot propagate, they only
  marginally contribute to the total current. Hence, speaking of a
  metallization of the MoS$_2$ monolayer does not appear justified
  when the involved electrons are not mobile \cite{iedm_contacts}. 
The large excess electron density in MoS$_2$ has nevertheless two
important consequences: (i) it pushes up the electrostatic potential
energy by $\Phi_P$, which acts as a barrier and blocks the transfer of
electrons from the Ti electrode to the MoS$_2$ layer in the overlap
region (high transfer resistance) and (ii) it partly pins the Fermi
level of the 2-D channel by screening the influence of the back gate
voltage. 

The situation is radically different in the presence of an interfacial
TiO$_2$ layer. The oxide prevents the penetration of the Ti
wavefunctions into MoS$_2$ and even depletes the electron population
in the overlap region, under flat band conditions. Due to this, the
MoS$_2$ conduction band can be readily modulated by the back gate
voltage and pushed below the metal Fermi level, thereby making itself
accessible for electrons tunneling through the oxide. This lowers the
transfer resistance from Ti to MoS$_2$. The usage of such Fermi level
de-pinning layer has been recently demonstrated for Co/h-BN electrodes
\cite{kim2}. 

\begin{figure}[t]
 \includegraphics[width=\linewidth]{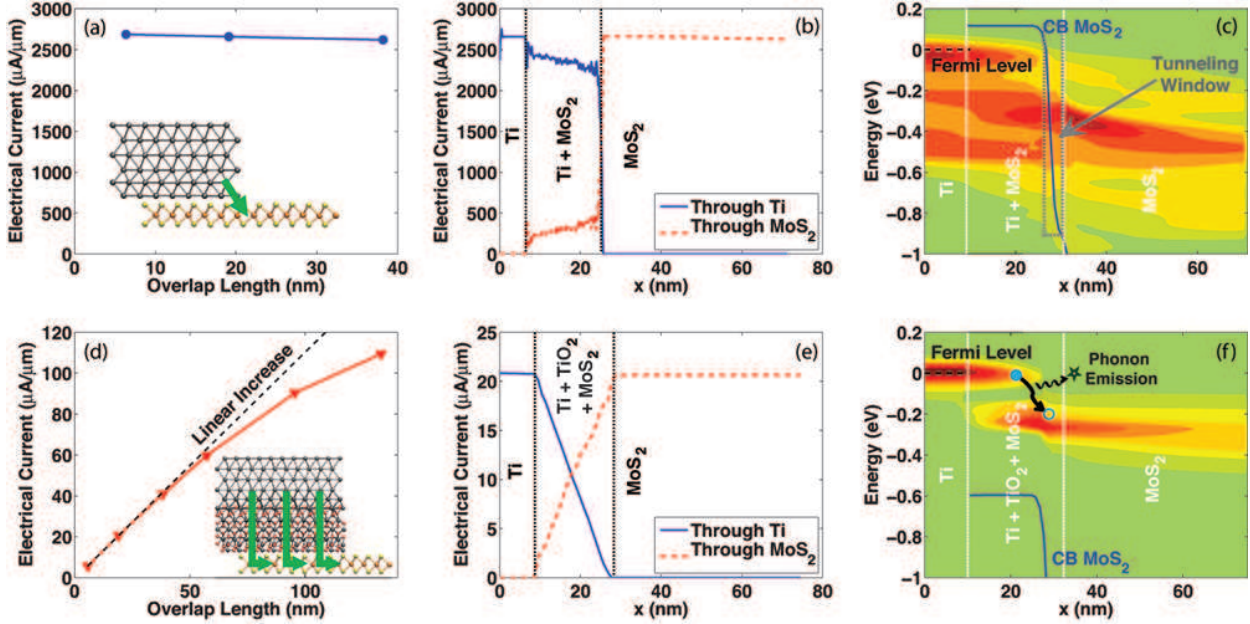}
  \caption{(a) Current flowing through the Ti-MoS$_2$ contacts in
    Fig.~\ref{fig:fig3} at $V_{gs}$=2 V and $V_{ds}$=2 V as
    a function of the overlap length $L_{overlap}$. The inset shows
    that the current (green arrow) is transferred from Ti to MoS$_2$
    at the contact edge. (b) Spatially resolved current along the
    $x$-axis in the Ti and MoS$_2$ regions of the structure with
    $L_{overlap}$=19.7 nm. An abrupt transfer from Ti to
    MoS$_2$ can be observed at the contact edge. (c) Spectral current
    distribution corresponding to sub-plot (b). Red indicates high
    current concentrations, green zero current. The blue line refers
    to the MoS$_2$ conduction band edge, the dashed line to the Fermi
    level, and the dashed rectangle, the electron tunneling
    window. (d-f) Same as (a-c), but for the Ti-TiO$_2$-MoS$_2$ 
    contacts. Since the current depends on the metal/MoS$_2$ overlap
    length, the whole process becomes area- and no more
    edge-dependent. This can be best seen by the gradual current
    transfer between Ti and MoS$_2$ in (e).}
  \label{fig:fig4}
\end{figure}

Using these electrostatic potentials, the current flowing through the
assembled Ti-contacted MoS$_2$ structures was computed at a back gate
voltage $V_{gs}$=2 V and a source-to-drain voltage $V_{ds}$=2 V
measured between the Ti region on the left (``source'') and the
MoS$_2$ single-layer on the right (``drain''). Figure \ref{fig:fig4}
unveils the behavior of the electrical current as a function of the
(i) metal-semiconductor overlap length $L_{overlap}$, (ii) region
where it flows (Ti or MoS$_2$), and (iii) energy at which it is
carried. It clearly appears that when Ti and MoS$_2$ are in direct
contact, the current becomes independent of $L_{overlap}$ and stays
constant (Fig.~\ref{fig:fig4}(a)). There is almost no current
exchange between the top electrode and the bottom 2-D material because
of the blocking potential barrier discussed above, except at the edge
of the Ti/MoS$_2$ interface (Fig.~\ref{fig:fig4}(b)) where the
combined effect of $V_{gs}$ and $V_{ds}$ pushes down the MoS$_2$
conduction band. This opens up a tunneling window for the electrons
situated in the Ti electrode, as can be seen in Fig.~\ref{fig:fig4}(c)
\cite{iedm_contacts}, activating a near-edge injection process whose
efficiency hinges on the characteristic (or screening) length
$\lambda_c$ of the contact \cite{app}. Here, due to the ultra-thin
bottom HfO$_2$ layer, $\lambda_c$ does not exceed 5 nm, but in
real devices, it is typically much longer.

The presence of an interfacial TiO$_2$ layer completely changes the
transfer mechanism. The current through the junction linearly
increases with the overlap length $L_{overlap}$ before slowly
saturating, as reported in Fig.~\ref{fig:fig4}(d). By extrapolating
this plot, a transfer length $L_T\sim$150 nm can be estimated. Figure
\ref{fig:fig4}(e) reveals that the electrical current gradually enters
the MoS$_2$ monolayer from the top metallic contact, over the entire
overlap region, which is a clear signature of an area-dependent
injection process: the longer the contact, the higher the current
magnitude. Although the quantitative $I_d$ values should be taken with
precaution because of the applied modeling approximations, in particular
the consideration of one single $k_z$ point, the performance of both
contact configurations can still be compared to each other. First, it
can be observed in Fig.~\ref{fig:fig4}(c) that without TiO$_2$, the
current distribution remains fairly homogeneous in the pure Ti and
Ti-MoS$_2$ overlap regions before losing energy when entering the
MoS$_2$-only extension due to phonon emission. When a TiO$_2$ layer is
inserted, phonon-assisted tunneling dominates the Ti-to-MoS$_2$
electron transfer, as indicated by the current distribution in
Fig.~\ref{fig:fig4}(f). This trend is confirmed when looking at the
ballistic current, which is 2.5 times lower than the one with
electron-phonon scattering.

From Fig.~\ref{fig:fig4}, it is also apparent that the total injected 
current is larger without the TiO$_2$ oxide layer despite the high
Ti-MoS$_2$ transfer resistance caused by the interface potential
barrier in the overlap region. This can be explained on one hand by
the fact that electrons following the diagonal path shown in the inset
of Fig.~\ref{fig:fig4}(a) move through a typical Schottky contact whose
triangular shape can be modulated by the applied gate-to-source
voltage. In this case, a larger characteristic length $\lambda_c$, as
encountered in real devices, would significantly decrease $I_d$ by
making the tunneling distance from Ti to MoS$_2$ longer. On the other
hand, when TiO$_2$ is present, electrons must tunnel through this
oxide to reach the 2-D channel, which reduces the transfer
probability. A thinner, more transparent, interfacial layer,
e.g. h-BN, could enhance the current magnitude.

It remains to put our work into perspective with literature. As
already mentioned above, the metallization of MoS$_2$ below Ti
contacts proposed in Ref.~\cite{Kang} does not support an
area-dependence of the electron injection. On the contrary, it pins the
Fermi level and deteriorates the modulation of the electrostatic
potential in the overlap region. The experimental data of
Refs.~\cite{large_transfer_length} and \cite{English} agree with our
results if we assume that there is a TiO$_2$ oxide layer between Ti
and MoS$_2$ in the former case, whereas the interface is devoid of any
oxide layer in the latter due to the higher vacuum deposition
conditions. The modeling-based Ref.~\cite{iedm_contacts} postulated a
near-edge injection process because a pure Ti-MoS$_2$ stack
(without TiO$_2$) was simulated. The same was found in
Ref.~\cite{meff_contact} for top contacts on a single-layer of
MoS$_2$. However, the authors of this paper noticed a transition from
an edge- to an area-dependent process as the MoS$_2$ thickness
was increased, implying that any interfacial layer could be beneficial as
long as it attenuates the penetration of the metal wavefunction into the
band gap of the bottom semiconductor layer(s).

In conclusion, we have used \textit{ab initio} simulations to
demonstrate that the injection of electrons from a top metallic
contact into an underlying 2-D material can occur either at the edge
or through the metal-semiconductor overlap area, depending on the
presence or not of an interfacial layer. In this paper, Ti electrodes
deposited on a MoS$_2$ layer, with and without an intermediate
TiO$_2$ oxide, have served as an example to illustrate the physics at
play. This finding can in principle be generalized to any blocking
layer placed at the  interface between a top contact and a 2-D monolayer,
intentionally or not. Such a layer can hinder the penetration of the
wavefunction originating from the metal into the band gap of the
semiconductor, thus enabling an area-dependent transfer process. It
can be envisioned that by engineering the properties of the
interfacial layer the contact resistance of FETs based on 2-D
semiconductors could be reduced, for example by selecting a material
with a conduction band edge well-aligned with that of the 2-D
crystal. Mobile electrons could then be directly injected into the
transistor channel, without tunneling. At the same time, the charges
pinning the Fermi level would still be stopped by the interfacial layer.

\begin{acknowledgement}
This research was supported by ETH Z\"urich (grant ETH-32 15-1) and
the Swiss National Science Foundation und grant no. 200021\_165841 and
under the NCCR MARVEL. We acknowledge PRACE for awarding us access to
Piz Daint at CSCS under Project pr28 and CSCS for Project s876.
\end{acknowledgement}

\begin{suppinfo}
Detailed description of the modeling approach and of the applied
approximations. 
\end{suppinfo}

\bibliography{nano}

\end{document}